# An Intelligent Vice Cluster Head Election Protocol in WSN

**Ghassan Samara, Mohammad A. Hassan, and Yahya Zayed**

Computer Science Department, Zarqa University, Zarqa- Jordan
e-mail: gsamarah@zu.edu.jo
Computer Science Department, Zarqa University, Zarqa- Jordan
e-mail: mohdzita@zu.edu.jo
Computer Science Department, Zarqa University, Zarqa- Jordan
e-mail: Yahyazayed88@gmail.com

**Abstract**

***Wireless sensor networks (WSNs) has a practical ability to link a set of sensors to build a wireless network that can be accessed remotely; this technology has become increasingly popular in recent years. Wi-Fi-enabled sensor networks (WSNs) are used to gather information from the environment in which the network operates. Many obstacles prevent wireless sensor networks from being used in a wide range of fields. This includes maintaining network stability and extending network life.***

***In a wireless network, sensors are the most essential component. Sensors are powered by a battery that has a finite amount of power. The battery is prone to power loss, and the sensor is therefore rendered inoperative as a result. In addition, the growing number of sensor nodes off-site affects the network's stability. The transmission and reception of information between the sensors and the base consumes the most energy in the sensor.***

***An Intelligent Vice Cluster Head Selection Protocol is proposed in this study (IVC LEACH). In order to achieve the best performance with the least amount of energy consumption, the proposed hierarchical protocol relies on a fuzzy logic algorithm using four parameters to calculate the value of each node in the network and divides them into three hierarchical levels based on their value. This improves network efficiency and reliability while extending network life by 50 percent more than the original Low Energy Adaptive Clustering Hierarchy protocol.***

# 1 Introduction

In scientific research, wireless sensor networks (WSNs) play a critical role. A wireless sensor is a gadget with a limited number of power sources that collects information from the environment. The field of wireless sensor networks (WSNs) is experiencing rapid growth due to the benefits it provides in a variety of fields and the human desire to know as much as possible about the environment [1]. To collect data from the environment and send it back to a base for processing or saving in a database for later use, wireless sensors are employed in a wide variety of industries, including weather forecasting and natural phenomena research as well as tracking bird migration and marine navigation [2 and 3].

For researchers and developers, the network's sensors' limited resources provide a barrier because of their limited capabilities. In order to work, gather data, and send and receive them, sensors rely on batteries as a finite energy source. (WSN) must be stable if it is to provide the most significant advantage to its users. For this reason, energy-saving measures need to continue to be refined and improved [4, 5, 6, 7, and 8].

On exhaustion of battery capacity, the network loses this sensor, and the greater number of sensor losses causes a rise in energy consumption and loss of more sensors. As a result of increased distances between sensors, there are a more enormous energy drain and more sensor losses [9, 10, 11. 12]. This has a negative impact on the network's stability and the quality of the data collected, resulting in a false impression of the surrounding environment [13, 14, 15, and 16]. For wireless sensor networks (WSNs) to reduce their energy consumption, they must improve the operation of sensor components that have two portions. As the name implies, it consists of two parts: hardware and software [17, 18, and 19].

As a result of the rapid growth of wireless sensor networks, sensors process the received data to keep track of particular events or make decisions for the network's users, such as how to maintain the network's stability and extend its operating life. As a result, this network has become a major source of anxiety [20, 21, and 22]. Therefore, it is necessary to find ways to reduce the power consumption of these sensors because they are powered by small batteries [23].

When the sensor power runs out, the network loses this sensor, and as the network loses more sensors, the burden on the remaining sensors grows. When sensors are further apart, energy consumption increases and a greater number of sensors are lost more quickly, resulting in higher energy costs. In order to reduce energy consumption, protocols must be developed to govern the ways and techniques of work, as well as the equal distribution of jobs among sensors, and to limit the energy depletion of some sensors compared to others [24].

Transmission and receiving are the most battery-intensive processes performed by the nodes. The transmissions and receptions are regulated by the communication protocols that the network relies on. Communications protocols will become more efficient as they are developed, which will reduce energy consumption [25, 26, 27, 28].

This paper aims to propose a protocol relies on artificial intelligence algorithm deployment to make a communication between nodes and the base through cluster head (CH) and CH Vice election, CHsec, and CHsec Vice election, To achieve load balancing in power consumption during message transmission between the nodes and the base furthermore, increase network stability by improving its performance, and extending network service lifetime.

## 2 Related Work

As a result of the previous period's research, it's anticipated that this field will continue to be researched for a long time to come as well. Several solutions have been proposed by researchers in an effort to determine the best approach to prolong the life of the network while maintaining its stability and sustainability. As WSN becomes more efficient, it will be used in more applications. Several prior studies have looked at strategies to reduce energy consumption in WSN networks, and we'll take a look at a few of them in this section.

Low-Energy Adaptive Clustering Hierarchy The (LEACH) protocol was the first protocol based on electing a cluster head (CH). The LEACH protocol selects CH at random, and BS has no idea how much energy is left in the node or how far it is from the base station. Everyone in the network has an equal chance of becoming CH, regardless of their position in the network [29]. Choosing a low-power node increases the likelihood of it dying sooner, resulting in network instability. It is equally likely that a CH node will be selected in the second round if it completes its task in Round 1. If it's set a second time, the node's chances of failing go up. They offered new modified LEACH methods to overcome these issues (see Fig. 1).

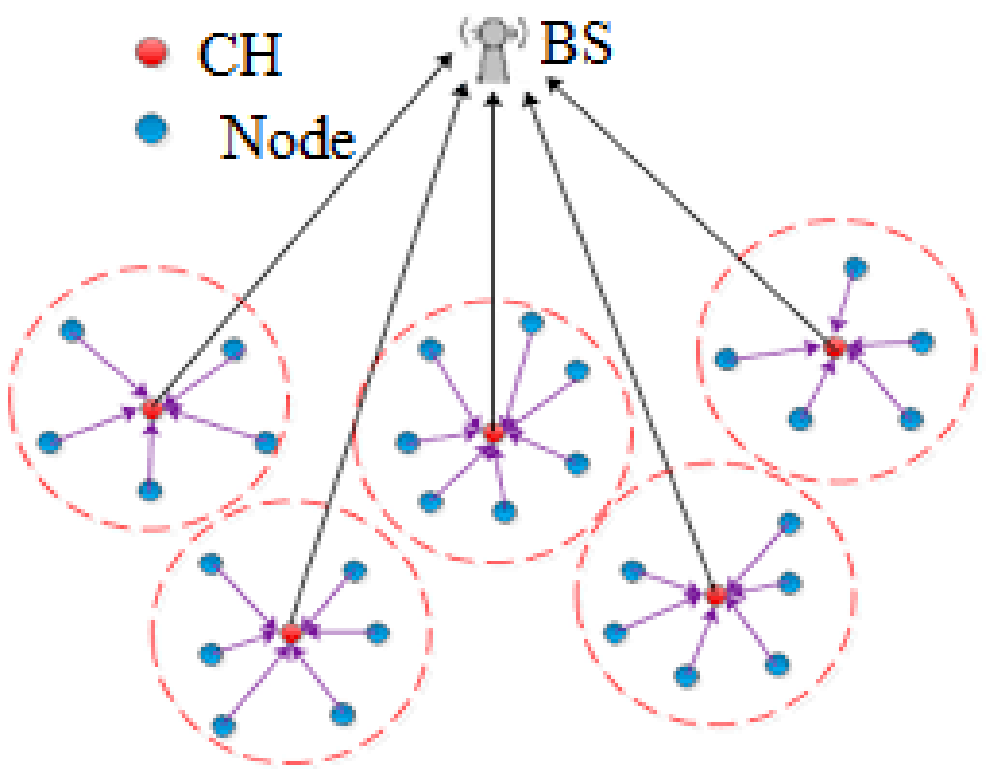

Fig. 1 LEACH protocol

In [30], authors proposed Enhanced Low Energy Adaptive Clustering Hierarchy (E-LEACH) introduced a new form for estimating sensor value in a node that has the same power level and the likelihood of becoming a cluster head is now available. Energy levels in the nodes will fluctuate after each round. To determine how much energy is left, the remaining energy levels are analyzed. "The Cluster Head" declares itself to be the node with the highest level of energy. In this way, the cluster head can be determined without a base station being required.

Low Energy Adaptive Clustering Hierarchy Centralized (LEACH-C) Clusters are formed by using a centralized approach in this protocol. It's not necessary to regroup after completing the cluster formation procedure. Cluster head nodes are rotated exclusively within their groups, and the normal nodes remain unchanged from LEACH [31]. There is no need to regroup once the cluster number has been identified and maintenance is performed throughout the network. After clustering, nodes cannot be added or removed, and the behavior of nodes after death cannot be regulated with this protocol.

Two-Levels Hierarchy for Low-Energy Adaptive Clustering Hierarchy (TL-LEACH) Two degrees of authority has been proposed in this protocol, the major cluster head and secondary cluster head [32]. To accomplish the task, the secondary cluster head collects data from all of the nodes and then delivers it to the base.

In [33], Base-Station Controlled Dynamic Clustering Protocol (BCDCP) protocol was proposed to solve the base's inability to determine node's position or remaining energy. A base has full decision-making authority under this protocol. Nodes are equipped with a GPS system that transmits their location and remaining energy to the base station via satellite. In this protocol, the usage of GPS and the transmission of information and choices from the base resulted in significant energy loss.

Authors proposed Vice-Low-Energy Adaptive Clustering Hierarchy (V-leach) protocol in [34], where cluster's nodes are divided into three categories to reduce power usage. CH-nodes are responsible for receiving and transmitting data to the base station, CHS-nodes, the Deputy Leader, and normal nodes that collect data and deliver it to the Leader CH. They are also responsible for receiving and sending data to the Deputy Leader. The protocol relies on CHS to take over all of CH's tasks in the event that CH fails. This protocol depends on three key parameters to determine if a node is CH, CHS, or normal: the distance between the base and the node, the node's maximum energy, and the node's minimum energy.

Advanced Low Energy Adaptive Clustering Hierarchy (A-LEACH) is proposed in [35]. During the hierarchical creation of the cluster, a new type of node called the CAG node was created. These are high-energy nodes that gather data from normal nodes in the cluster and deliver it to messaging portals or data collecting basins. This will lower the power consumption of cluster head nodes and save energy.

Multi-hop Routing with Low Energy Adaptive Clustering Hierarchy (MR-LEACH) is proposed in [36], by dividing the network into clusters, choosing which cluster head will be in charge of collecting and transmitting data, and then determining the optimal method to transfer and collect data between each cluster head, in order to save expenses in data transfer.

K-LEACH is proposed in [37]. This approach uses the k-medoids algorithm, which ensures that network power is divided across all active nodes in a manner similar to the original LEACH protocol. Once the nodes are sorted out, it begins to work on rearranging them. It is possible to locate CH between the active nodes by arranging the nodes. This is followed by a random selection of CHs based on the quantity of energy remaining when 50 percent of the network operations are completed. The K-LEACH concept increases network efficiency by not identifying CHs at random.

Improved-LEACH (I-LEACH) is proposed in [38]. The design includes four major variables such as the method of clustering, data aggregation, mobility, and scalability. The success of this process and the amount of residual energy relied on selecting to avoid adopting the same node twice in a row. But, at some point, the preceding CH will be the only choice when this task must be performed. Not using it would, according to the author's information, drain all nodes in the network one by one.

Authors in [39], adopting distances from the network's lifetime leads to improved network organization. The higher the energy consumption when sending and receiving data, which was evident with nodes having an increased distance from the base, they also ended up wasting resources.

ESO-LEACH is proposed in [40]. The researchers used ESO-LEACH to piece together their sensor nodes. To lessen the randomness of the ESO-LEACH method, the advanced nodes idea and upgraded rule set are employed to choose CHs. The Python-based ESO-LEACH concept demonstrates a significant increase in network longevity over LEACH. When compared to the present LEACH protocol, the ESO-LEACH network lifetime was twice as long.

Partitioned-Based Energy-Efficient–LEACH (PE-LEACH) is proposed in [41]. In this research, the writers' reliance on LEACH improvement helps increase the quality of their work. The authors have developed a new convention, partition-based clustering, which operates on the LEACH clustering method. A protocol used to strengthen energy-based fault tolerance technology (PE-LEACH). Additionally, they examined classifying the LEACH variations. Compared to previous LEACH incarnations, PE-LEACH has been studied using a combination of computing LEACH, E-LEACH, and LEACH (ESO-LEACH). They found that PE-LEACH outscored LEACH and E-LEACH independently while ESO-LEACH, its most competitive competitor, is doing well, too.

A Fuzzy Logic-Based Clustering Algorithm for WSN to Extend the Network Lifetime is proposed in [4]. In certain WSN studies, unclear methodologies have been employed to combine nodes. To help the network run longer, a fuzzy-based clustering strategy has been proposed. By choosing CHs who can send information to the mobile BS by using only well-chosen descriptors such as remaining battery power, BS mobility, and cluster centralization, the authors say it is possible to determine a super-CH (SCH) by doing so. The authors utilize the Mamdani rule to choose CHs in the Fuzz Engine. It was found that Leach's protocol extended network life by 25%. Methodology.

# 3. The IVC LEACH

To help save power in the nodes, the proposed protocol (IVC LEACH) organizes the network and divides it to reduce correspondence. This protocol uses five types of nodes in each group, with these classifications:

- CH Node (Cluster Head).
- CHv Node (Cluster Head Vice).
- CHsec Node (Cluster Head second).
- CHsecv Node (Cluster Head second Vice).
- Normal Node.

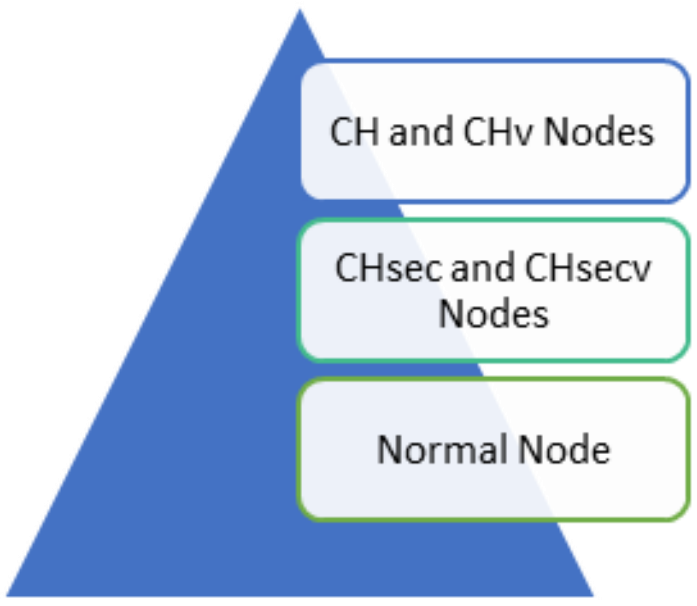

Fig. 2: Hierarchical Structure of Nodes in the Network.

At the top of the hierarchy are the CH node and its deputy (CHv node). In the second level, the CHsec node and its deputy CHsecv are placed. In the third level, the rest of the nodes are placed in the group, which are the normal nodes that collect data from the surrounding environment.

The fundamental purpose of this hierarchy is to increase the amount of available power in the network by reducing correspondence between nodes.

Once the nodes have been separated, the process of calculating the value of each node begins, with the value of each node being derived from three essential variables:

- The power value remaining in the node.
- The distance between the node and the BS.
- The space between the node and the other nodes.

In order to calculate this value in the proposed protocol (IVC LEACH), the Fuzzy-Logic algorithm was adopted.

After determining the value of each node, the nodes are grouped by value. The CHsec node is like a basket where data from other nodes is gathered, and the CH node is used to communicate with the BS.

CH node is used to communicate with the BS.

In the case of a CH failure, the CHv node is activated, and if the CHsec fails, the CHsecv node is engaged.

The steps taken to construct this protocol are shown in Fig. 3.

The IVC protocol is made up of two phases: configuring and steady-state. We shall examine these phases in more depth later.

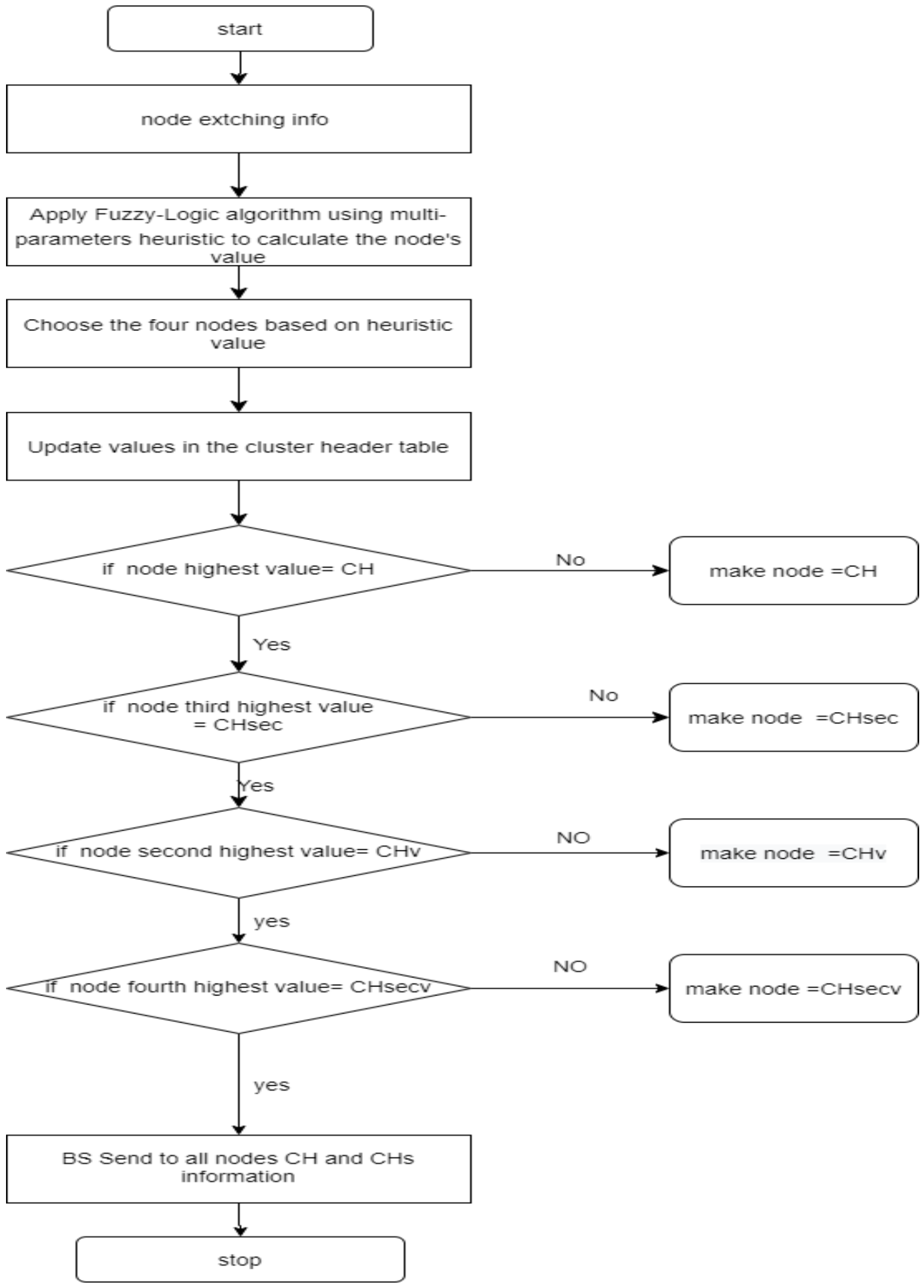

Fig. 3: The Methodology.

### *3.1 CONFIGURATION PHASE*

To begin this process, you must first identify the geographic area where a WSN will be created. In order to cover the given area, all sensor nodes must be deployed at random. It's important to note that the random distribution of nodes in the defined area causes a number of faults that might have a negative impact on a network's performance, such as damaged nodes that are present or nodes that are unstable. After node deployment, the IVC LEACH protocol activates the network to be used. Once the process of setting up nodes in the network is done, each node sends a message to BS, which has three pieces of information: the kind of the node, the ID of the node, and the location of the node:

- The remaining energy in the node.
- The location of the node.
- Node ID.

When the BS receives messages from the nodes, it analyzes this data to Fig. out the network's topology and divide the nodes into groups based on their positions. Another aspect of the Clustering Algorithm is the Fuzzy-Logic algorithm, which calculates the heuristic value of each node in a cluster using four parameters: However, this information includes the node's remaining energy, distance from BS, location in the network, and whether or not the node was utilized as a CH in the previous round. This information is used to determine the heuristic value of each node in equation (1).

$$(R+D+C)*p = value \qquad (1)$$

Where:

R= remaining energy of the original energy node%.

D= distance from BS.

C= Central.

P= CH in the previous round.

Fig. 4 illustrates the amount of energy (R) each node has left according to the Fuzzy-Logic algorithm. And these energy levels are divided out over three possible categories: low, medium, and high.

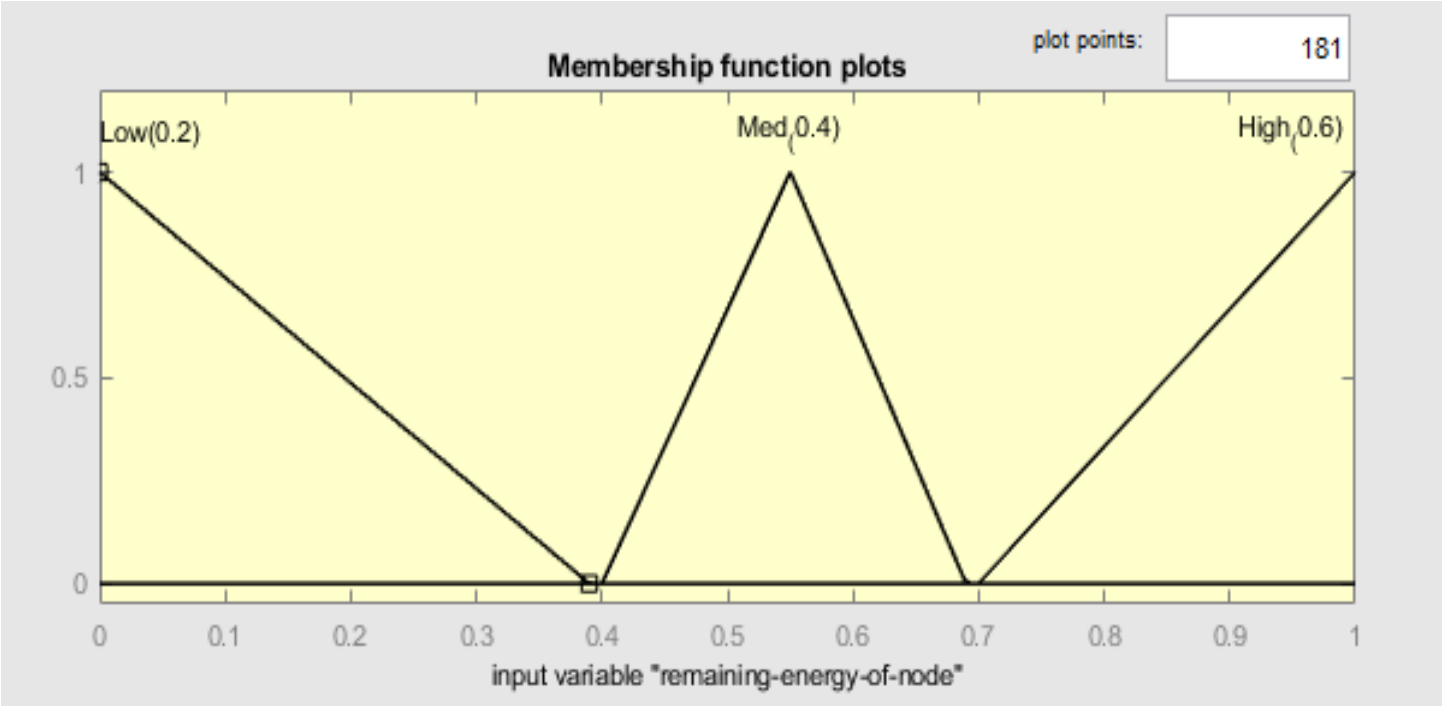

Fig. 4: Remaining Energy Of Node

Table 1 demonstrates the residual energy in each node, distributed in three levels: low (0.2), medium (0.4), and high (0.6).

Table 1 Remaining Energy of Node

| Remaining Energy Of The Original Energy Node% | Value |
|---|---|
| 0%-39% | Low(0.2) |
| 40%-69% | Med (0.4) |
| 70%-100% | High (0.6) |

Fig. 5 shows the distance from BS to the end of zone (D) in each node using the Fuzzy-Logic algorithm, and this distance is distributed in three values, close, med, and far.

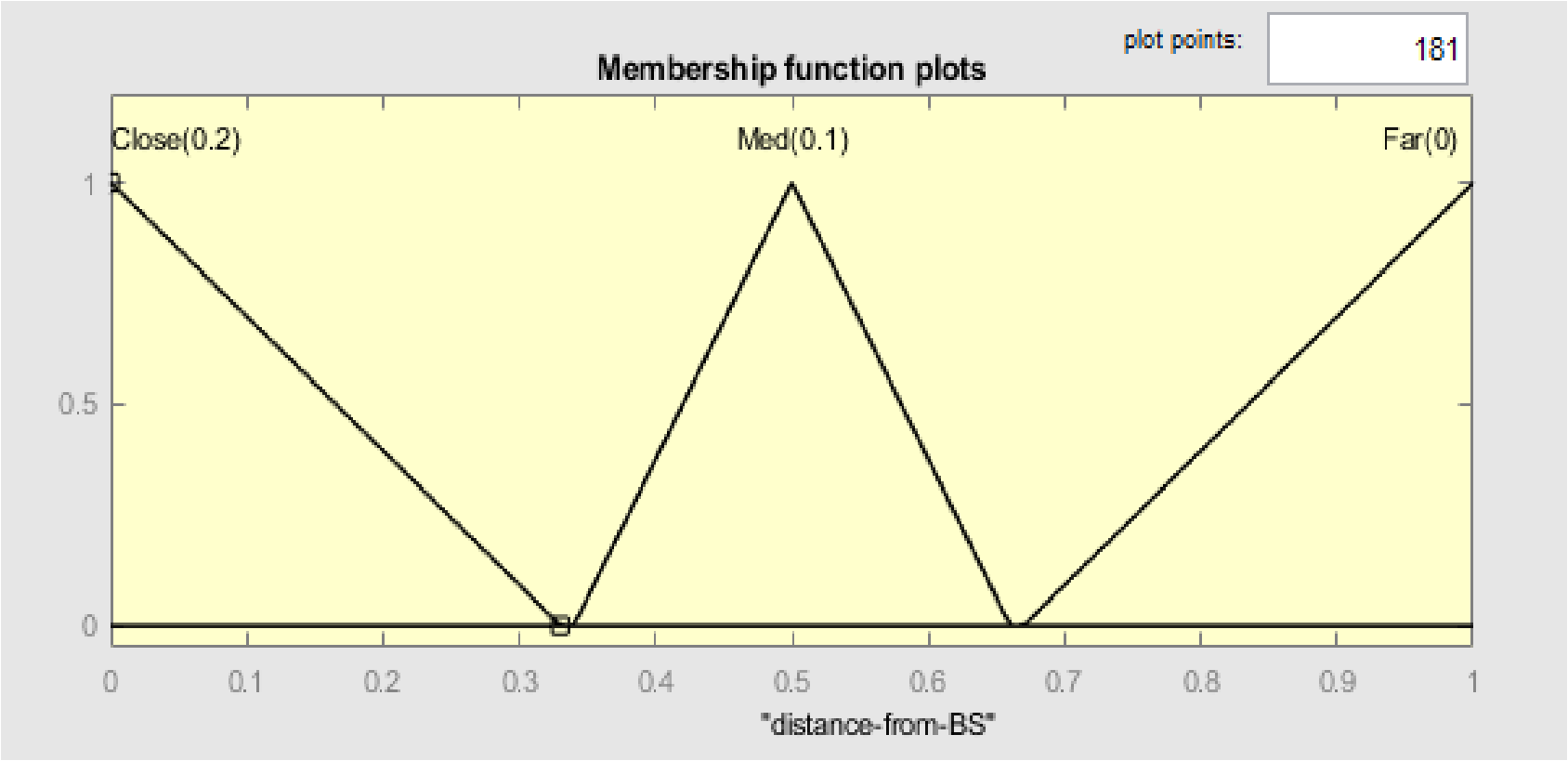

Fig. 5: Distance from BS

The distance from BS to the end of the D zone is shown in table 2, which was calculated with the Fuzzy-Logic algorithm, and the results are divided into the three ranges: close (0.2), med (0.1), and distant (0).

Table 2: Distance from BS Value

| distance from BS to the end of zone | Value |
|---|---|
| 0%-33% | Close(0.2) |
| 34%-66% | Med(0.1) |
| 67%-100% | Far(0) |

Fig. 6 demonstrates the use of Fuzzy-Logic in the Central node, and this node is able to exist in two states: in the center and in the edge.

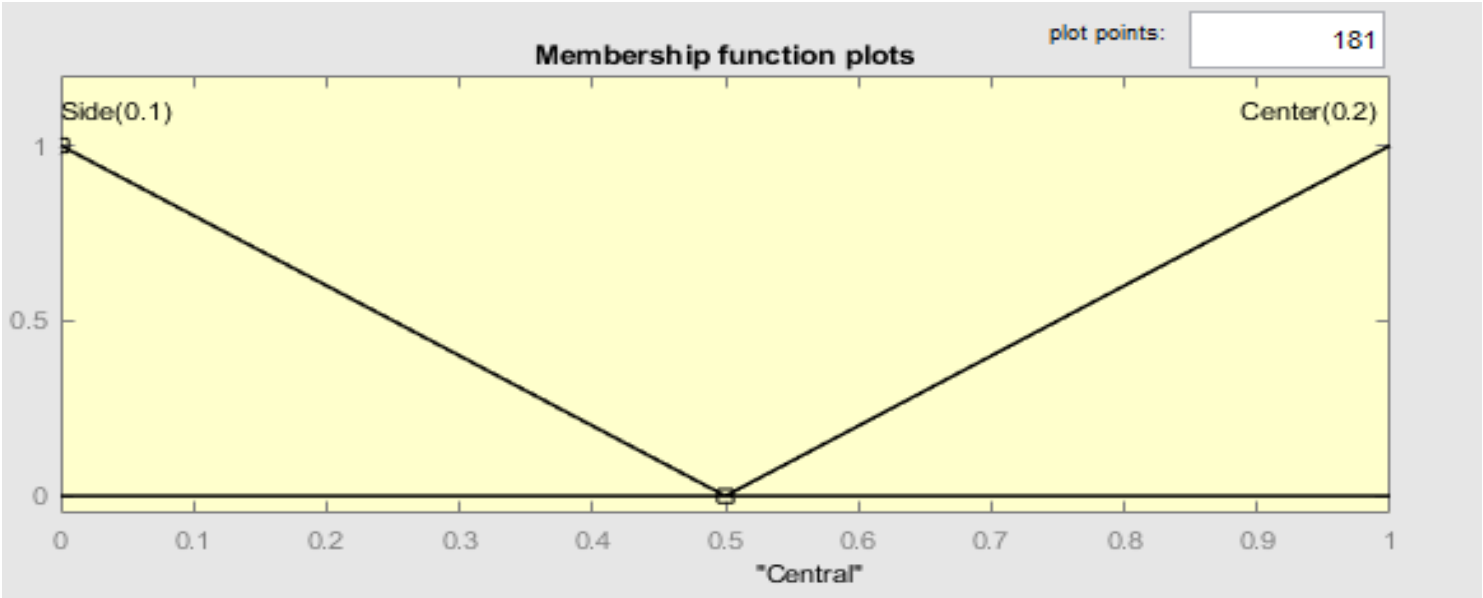

Fig. 6: Central

Table 3 illustrates how the C values in each node were calculated using Fuzzy-Logic, and this C is distributed into two values, a side cluster side (0.1) and a center cluster center (0.2).

Table 3: Central

| Central | Value |
|---|---|
| Side of cluster | Side(0.1) |
| Center of cluster | Center(0.2) |

The image in Fig. 6 depicts the CH utilizing the Fuzzy-Logic algorithm in the previous round (P), and it is dispersed into two values, yes or no.

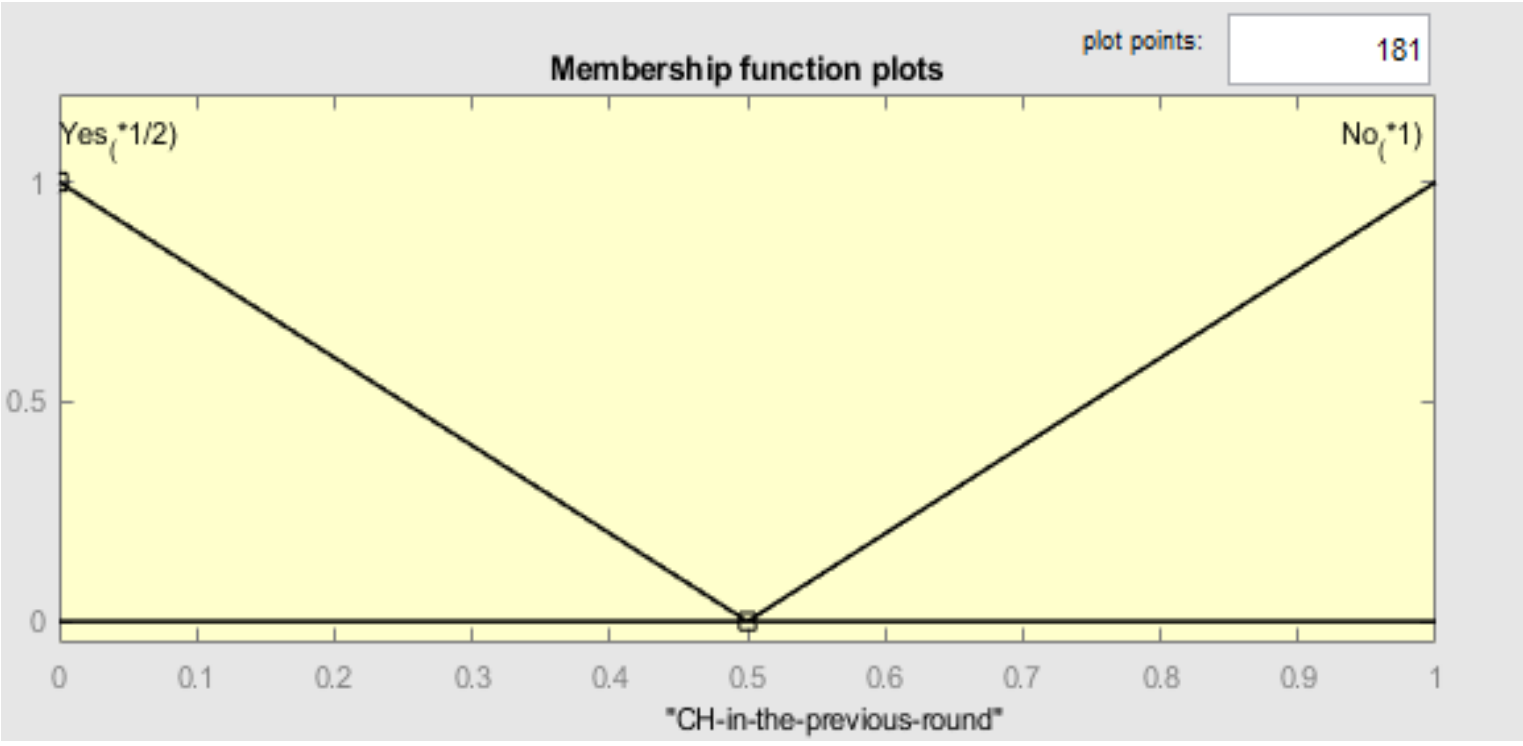

Fig. 7: CH in the previous round

As shown in Table 4, the Fuzzy-Logic algorithm determines if a node was CH in the previous round (P), and the results are either yes (the node was CH during the last round) or no (the node was not CH in the previous round).

Table 4 CH in the previous round

| CH in the Previous Round | Value |
|---|---|
| node was CH in the previous round | Yes (*1/2) |
| node was not CH in the previous round | No (*1) |

Fig. 8 displays the value of a node, which has four values, using the Fuzzy-Logic algorithm (remaining energy of the original energy node, distance from BS, Central, and CH in the previous round).

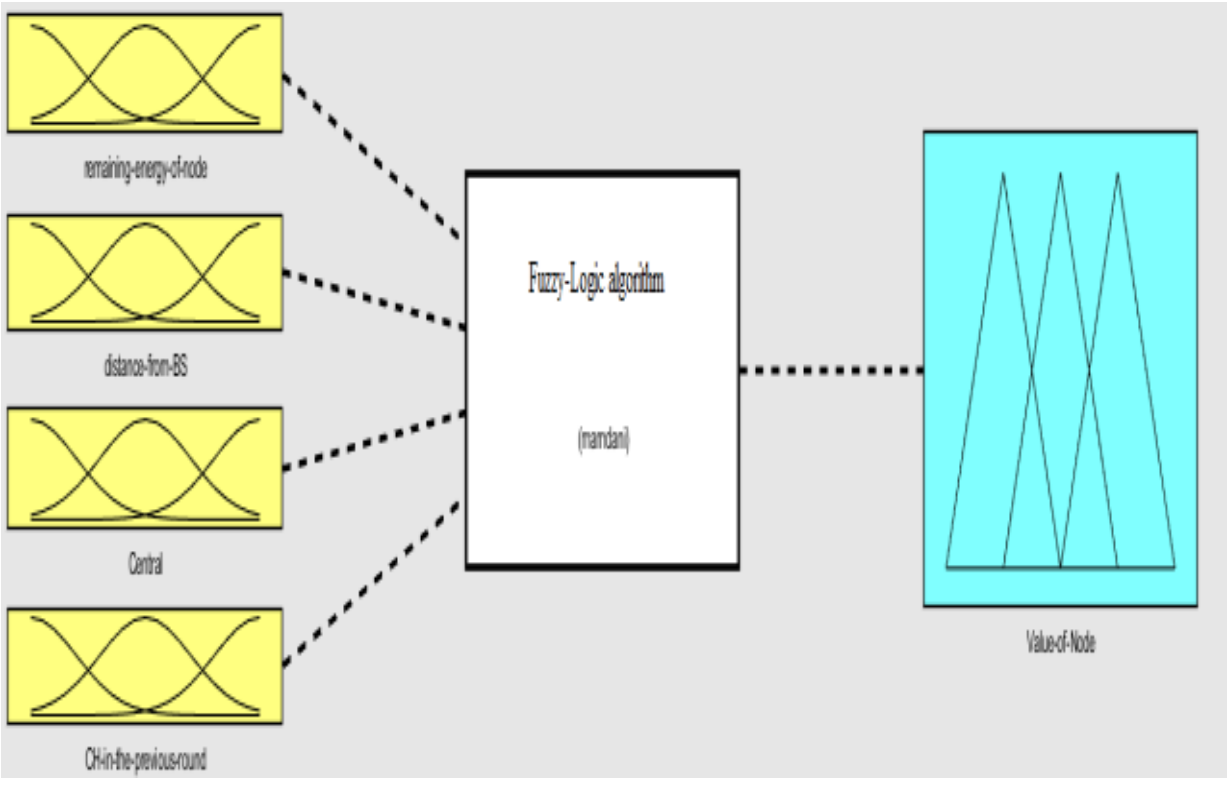

Fig. 8: Value of node

Using the Fuzzy-Logic technique, table 5 indicates the value of the node (remaining energy of the original energy node, distance from BS, Central, and CH in the previous round).

Table 5 Value of node1

| | Remaining Energy | Distance from BS | Central | CH in the Previous Round | Value of Node |
|---|---|---|---|---|---|
| 1 | Low(0.2) | Far(0) | Side(0.1) | Yes (*1/2) | 0.15 |
| 2 | Low(0.2) | Far(0) | Side(0.1) | No (*1) | 0.20 |
| 3 | Low(0.2) | Far(0) | Center(0.2) | Yes (*1/2) | 0.20 |
| 4 | Low(0.2) | Far(0) | Center(0.2) | No (*1) | 0.40 |
| 5 | Low(0.2) | Med(0.1) | Side(0.1) | Yes (*1/2) | 0.20 |
| 6 | Low(0.2) | Med(0.1) | Side(0.1) | No (*1) | 0.40 |
| 7 | Low(0.2) | Med(0.1) | Center(0.2) | Yes (*1/2) | 0.25 |
| 8 | Low(0.2) | Med(0.1) | Center(0.2) | No (*1) | 0.50 |
| 9 | Low(0.2) | Close(0.2) | Side(0.1) | Yes (*1/2) | 0.25 |
| 10 | Low(0.2) | Close(0.2) | Side(0.1) | No (*1) | 0.50 |
| 11 | Low(0.2) | Close(0.2) | Center(0.2) | Yes (*1/2) | 0.30 |
| 12 | Low(0.2) | Close(0.2) | Center(0.2) | No (*1) | 0.60 |
| 13 | Med (0.4) | Far(0) | Side(0.1) | Yes (*1/2) | 0.25 |
| 14 | Med (0.4) | Far(0) | Side(0.1) | No (*1) | 0.50 |
| 15 | Med (0.4) | Far(0) | Center(0.2) | Yes (*1/2) | 0.30 |
| 16 | Med (0.4) | Far(0) | Center(0.2) | No (*1) | 0.60 |
| 17 | Med (0.4) | Med(0.1) | Side(0.1) | Yes (*1/2) | 0.30 |
| 18 | Med (0.4) | Med(0.1) | Side(0.1) | No (*1) | 0.60 |
| 19 | Med (0.4) | Med(0.1) | Center(0.2) | Yes (*1/2) | 0.35 |
| 20 | Med (0.4) | Med(0.1) | Center(0.2) | No (*1) | 0.70 |
| 21 | Med (0.4) | Close(0.2) | Side(0.1) | Yes (*1/2) | 0.35 |
| 22 | Med (0.4) | Close(0.2) | Side(0.1) | No (*1) | 0.70 |
| 23 | Med (0.4) | Close(0.2) | Center(0.2) | Yes (*1/2) | 0.40 |
| 24 | Med (0.4) | Close(0.2) | Center(0.2) | No (*1) | 0.80 |
| 25 | High (0.6) | Far(0) | Side(0.1) | Yes (*1/2) | 0.35 |

| | Remaining Energy | Distance from BS | Central | CH in the Previous Round | Value of Node |
|---|---|---|---|---|---|
| 26 | High (0.6) | Far(0) | Side(0.1) | No (*1) | 0.70 |
| 27 | High (0.6) | Far(0) | Center(0.2) | Yes (*1/2) | 0.40 |
| 28 | High (0.6) | Far(0) | Center(0.2) | No (*1) | 0.80 |
| 29 | High (0.6) | Med(0.1) | Side(0.1) | Yes (*1/2) | 0.40 |
| 30 | High (0.6) | Med(0.1) | Side(0.1) | No (*1) | 0.80 |
| 31 | High (0.6) | Med(0.1) | Center(0.2) | Yes (*1/2) | 0.45 |
| 32 | High (0.6) | Med(0.1) | Center(0.2) | No (*1) | 0.90 |
| 33 | High (0.6) | Close(0.2) | Side(0.1) | Yes (*1/2) | 0.45 |
| 34 | High (0.6) | Close(0.2) | Side(0.1) | No (*1) | 0.90 |
| 35 | High (0.6) | Close(0.2) | Center(0.2) | Yes (*1/2) | 0.50 |
| 36 | High (0.6) | Close(0.2) | Center(0.2) | No (*1) | 1 |

# 4 Scenario

In this instance, we will highlight two nodes in Networks A and B. The residual energy in node A was 50% of the initial energy of the node, the node was at a medium distance from BS 45%, the node was on the side of the cluster, and the node was not CH in the previous round.

Node B had leftover energy of 30% in its node after being 45% from BS. It was in the middle of the cluster, and was designated CH in the previous round, as seen in Fig. 9 (scenario). The node value is based on the parameters of the nodes and their values.

-value of the node A= (0.4 + 0.1 + 0.1) * 1 = 0.60

-value of the node B= (0.2 + 0.1 + 0.2)*1/2=0.25

In this scenario node A gets a higher value than node B, and in this case, node A gets a greater chance than node B to be CH or CHsec or one of their Vice due to its greater value.

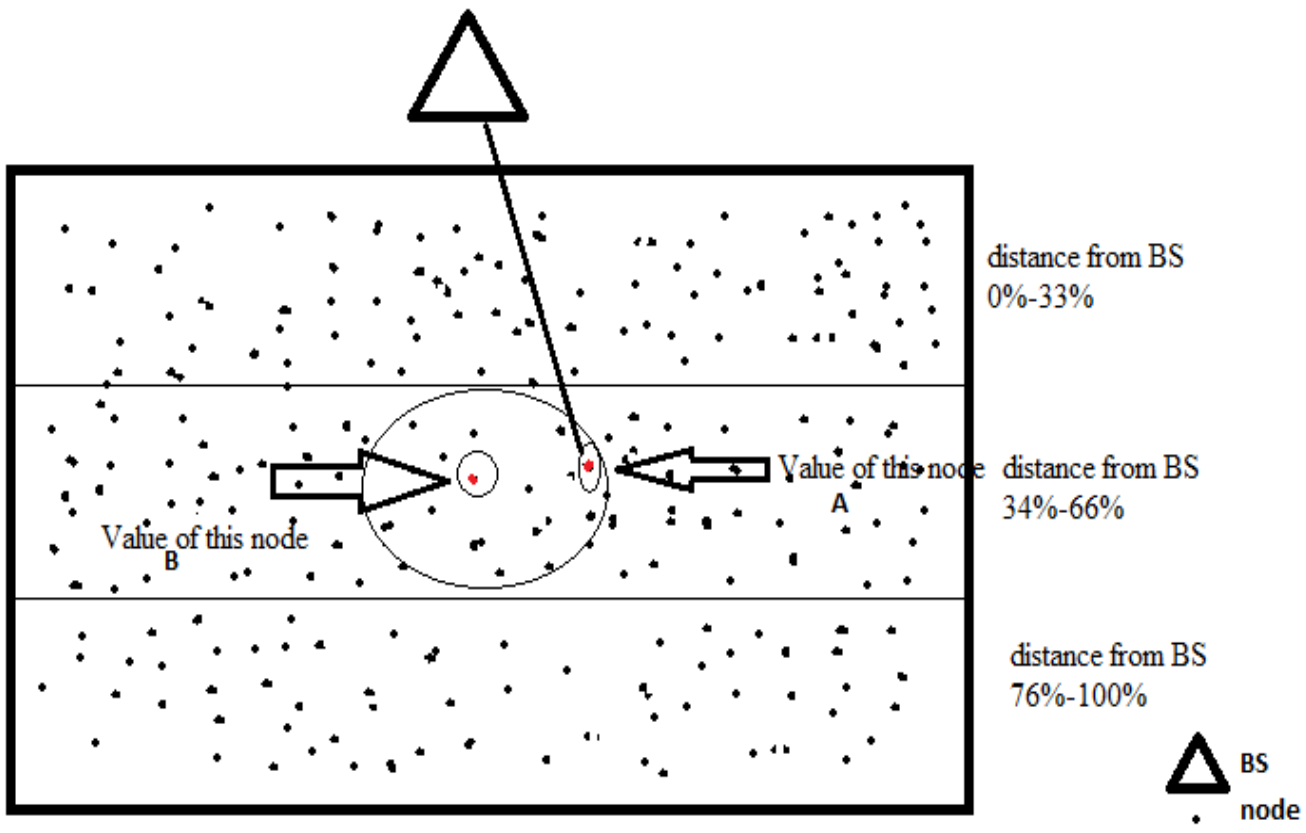

Fig. 9: Scenario

BS will first generate a table with all the node values for each cluster and then select the four best nodes in each cluster to create the CH table. There are four node entries in the CH table from each cluster, which is shown in the following table:

Table 6 The CH

| | |
|---|---|
| 1 | The CH node with the best value. |
| 2 | The CHsec with the second best value. |
| 3 | The CHv with the third best value. |
| 4 | The CHsecv the node with the fourth best value. |

Prior to this, the concept of calculating the number of clusters based on the number of cluster heads must be determined, or it will be fixed.

After partitioning the network into clusters and specifying CH, CHsec, CHv, and CHsecv for each cluster, BS broadcasts a message to all network nodes. And, once these procedures are completed, the network configuration phase will be completed, and the network will be operating.

STEADY-STATE PHASE

BS had already alerted the CHsec to this info. Afterwards, the CHsec sets a schedule for every node in the cluster. It sends a message to every node to let them know that they're transmitting on the schedule, telling them their allotted time to broadcast. This results in each CHsec node sending a message, which contains information gathered from its surroundings.

The message header contains the node ID and how much power is left in that node. After the end of each time slot, the node enters a "non-transmission" phase, generally known as "sleep," in which it waits for a second time slot to send more messages containing information gathered from the surrounding environment.

If the normal node does not receive a response from CHsec (after each time slot, CHsec sends a tiny message called live message containing its remaining energy to indicate that this message can still accept messages), the node begins sending to CHsecv, which immediately replaces the CHsec.

After the normal node's cycle of data collection is complete, the CHsec aggregates and passes data obtained from the normal node to the local CH. If CH does not answer (CH must send an AcK message), CHsec sends to CHv, who replaces CH immediately. The receiver node (either CH or CHv) will transfer the data to the BS after receiving it from CHsec. These data are divided into two parts: the first is the data collected from the environment, and the second is updated information about the current status of each sender node in the cluster (the node ID and the remaining energy in that node).

CH node, in turn, sends the information received from CHsec to BS, which utilizes the first portion of the information, which is information acquired from the environment surrounding the node after processing, in the application that employs WSN. BS uses the second part of the node's information to calculate new values for each node and update the values table, as well as the CH table, completing the first round. Following that, BS calculates the new nodes CH, CHsec, CHv, and CHsecv. The new information is sent by BS to the new CH nodes, kicking off the second round of information collecting from the network's nodes.

# 5 Simulation Results

Matlab R2017b was used to simulate the LEACH protocol and the suggested IVC-LEACH methodology. One hundred nodes in a network area was used to implement both protocols (100 m x 100 m). it is assumed that all nodes' starting Energy was equal, and we randomly dispersed the nodes in the network area using the simulator. BS is located at the top of the sensor network area, with coordinates (100, 50). The total number of election rounds used to generate a result is 2,500.

The two protocols are evaluated using four criteria listed below:

- Time consumed for the First Node to die
- Time consumed for the Last Node to die
- Lifetime rounds achieved by the network using both protocols
- Indicator of average network Energy loss during rounds.

Table 7 Parameters of simulation

| Parameters | Value |
|---|---|
| Network Area Size | (100 x 100) Meters |
| Number Of Nodes | 100 |
| Number Of Rounds | 2500 |
| BS Location | (100,50) |
| Initial Energy Of Node | 0.5 joules |

The network's coverage region is depicted in Fig. 10. The nodes were deployed at random and were used to demonstrate the outcomes of implementing the original Leach protocol and the suggested IVC-Leach protocol on the same network.

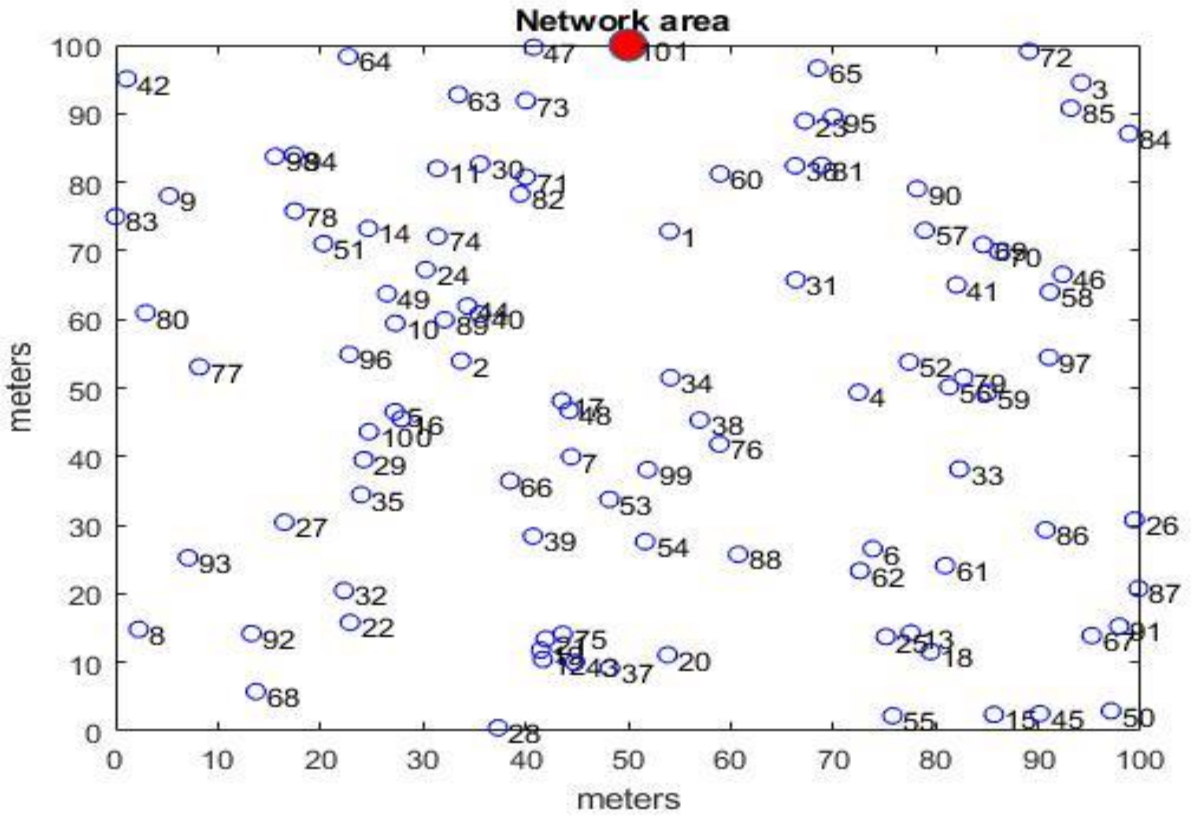

Fig. 10: Network Area

The implementation results of the original Leach protocol and the proposed IVC-Leach protocol are shown in Table 8. The results revealed that the Leach protocol outperformed the IVC-Leach protocol in the death of the first node due to the random selection of the cluster head. In contrast, the IVC-Leach protocol outperformed the Leach protocol in the death of the entire network node. Because the cluster head is determined by the calculated node value. The IVC-Leach protocol completed more network rounds than the Leach protocol.

Table 8 Implementation Result

| Parameters | LEACH protocol | IVC-LEACH protocol |
|---|---|---|
| First Node Dead | 964\round | 563\round |
| All Node Dead | 1491 \round | 2465\round |

Fig. 11 depicts a graph of the number of rounds obtained by the original Leach protocol and the suggested IVC-LEACH protocol in three scenarios: first, when the first node in the network died, second, when half of the nodes in the network died, and third, when all of the nodes in the network died. Through the Fig. that shows the superiority of the Leach protocol in the number of cycles it achieved until the death of the first node in the network, and this superiority is initially due to the random method of selecting CH, which leads to the energy consumption of the node randomly, then the IVC-LEACH protocol surpasses in the number of cycles that It achieves until the death of half of the nodes in the network. The reason for this is the distribution of energy consumption between CH and CHsec, as well as the manner of selecting them based on the node's value.

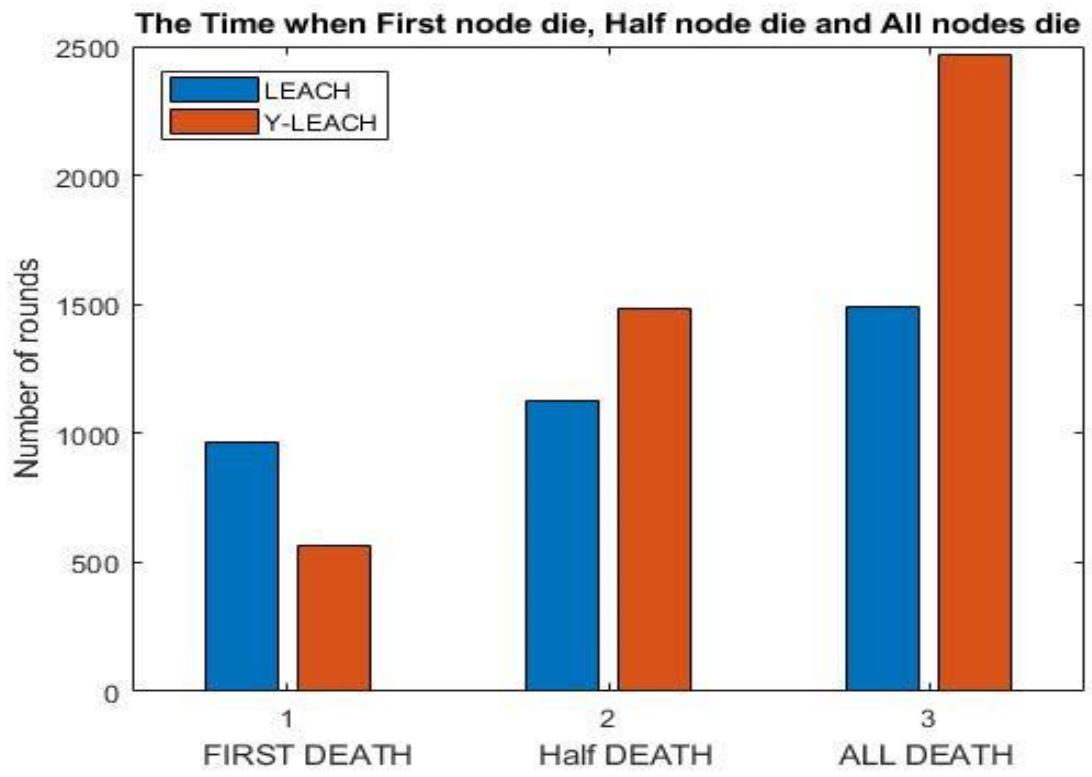

Fig. 11: Live Nodes Leach

According to Fig. 12, the IVC-Leach protocol lost the first node in round 480, while the original Leach protocol lost the first node in round 992. After that, the Leach protocol lost a huge number of nodes in the network unexpectedly within a few rounds, and the network lost all nodes in round 1550. The IVC-Leach protocol, on the other hand, continued to lose nodes gradually and did not experience a sudden drop in the number of network nodes, thus losing all nodes in the 2492 round, resulting in a higher number of cycles achieved by the network using the IVC-leach protocol than the protocol 50 percent original Leach.

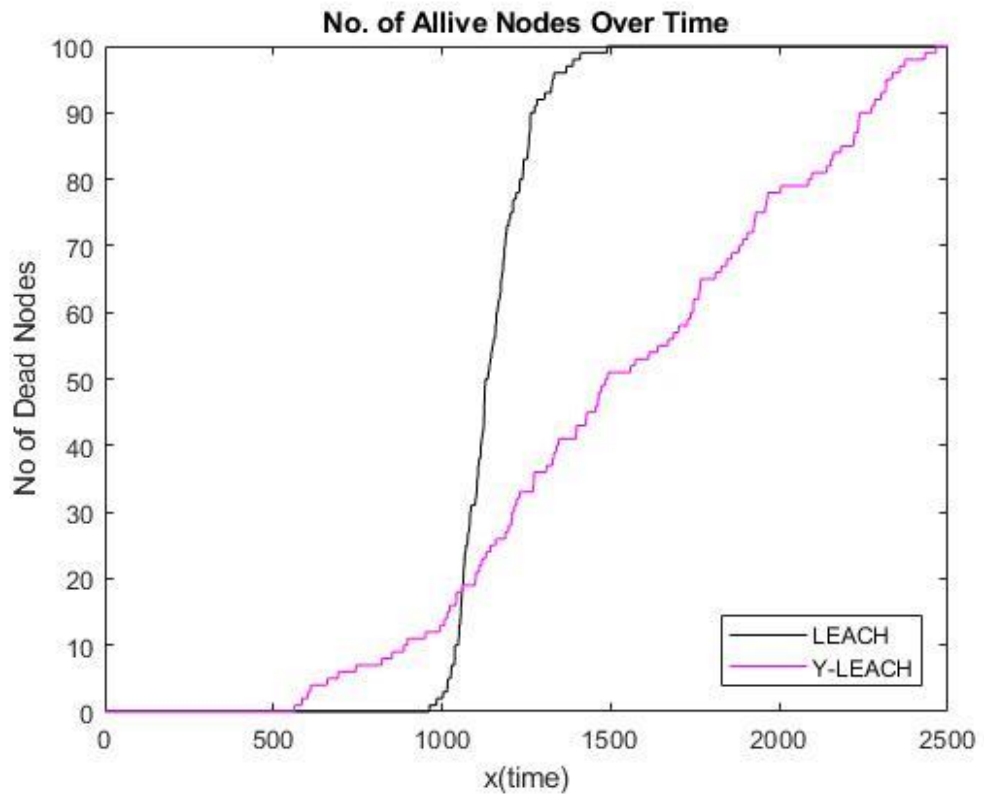

Fig. 12: Number Of Live Nodes

Fig. 13 depicts the number of nodes that die in the network during the network's duty rounds. Fig. 13 depicts the number of nodes that die suddenly in the Leach protocol due to random CH selection and the gradual decrease in the number of died nodes in the IVC-LEACH protocol due to the studied determination of CH and CHsec by each node value, resulting in the IVC-Leach protocol outperforming the Leach protocol by about 50%.

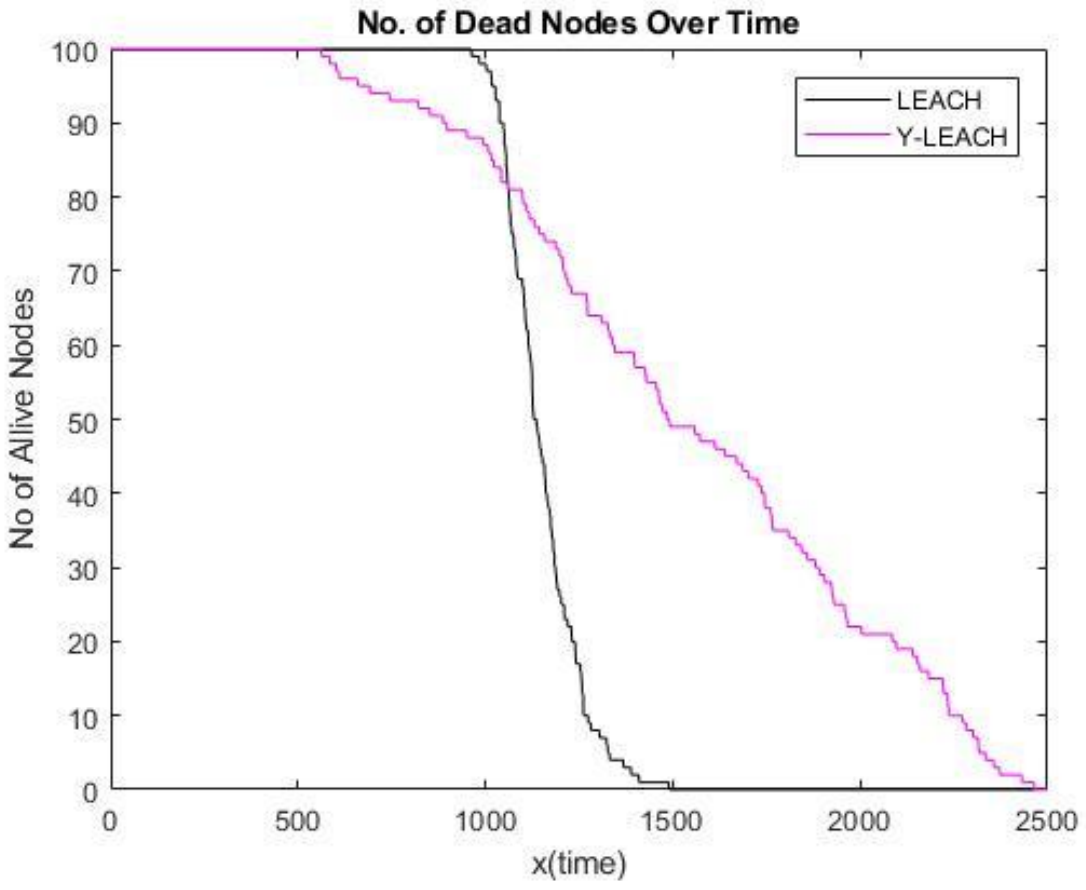

Fig. 13: Number Of Dead Nodes

Fig. 14 depicts the power level in the network while using the original Leach protocol and the IVC-LEACH protocol, and the IVC-LEACH protocol excels at conserving network power while working by distributing power consumption to network nodes rather than consuming power from a group of nodes in a larger way. From different nodes.

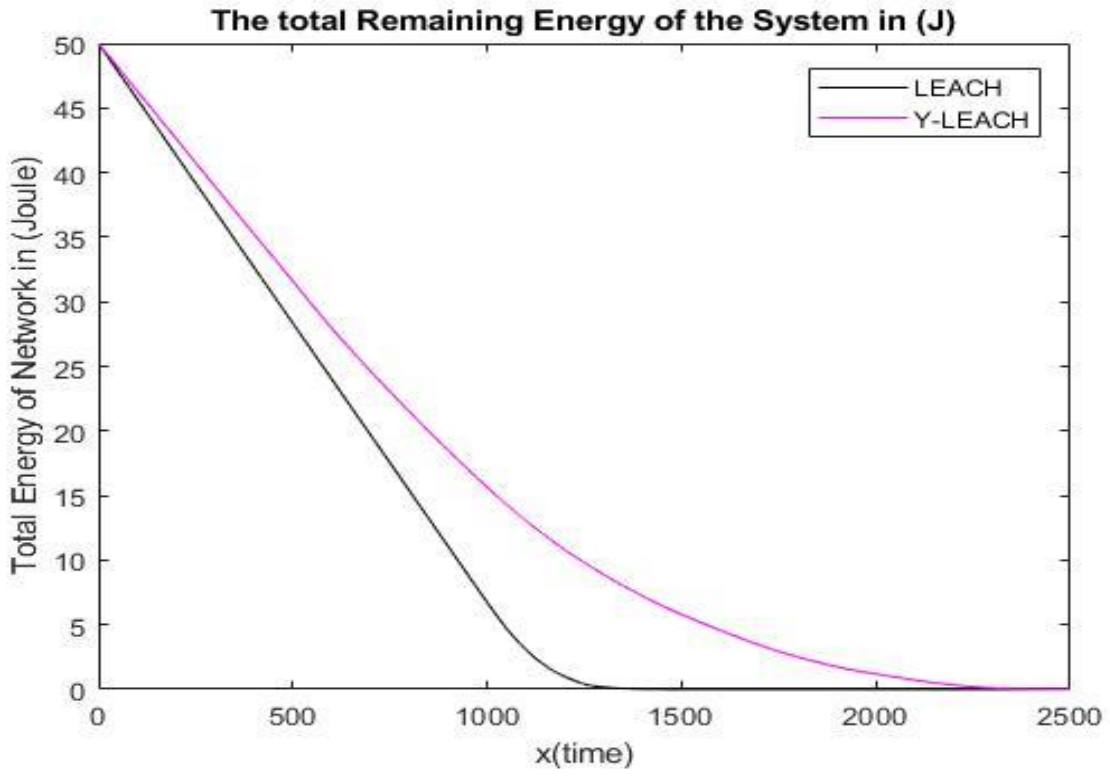

Fig. 14: Average Residual Energy

The implementation results show that the original Leach protocol loses the first node when the network reaches round 964, whereas the proposed IVC-LEACH protocol loses the first node in round 563, and the Leach protocol loses all nodes in round 1491, while the proposed protocol continues to operate and all nodes are lost in round 2465. It completed 1491 rounds on the network using the original Leach protocol and 2465 rounds using the proposed IVC-LEACH protocol. Table 8 implementation results demonstrate that the proposed IVC-LEACH protocol outperforms the original Leach protocol by 50% due to a significant difference in the number of rounds of work accomplished by the network utilizing IVC-LEACH increased network stability and network life.

# 6 Conclusion

Because of the importance of wireless sensor networks and their numerous applications in many bands, these networks and the challenges they may meet while working has been thoroughly researched in order to improve and raise their efficiency. After finishing the investigation of wireless sensor networks, this paper developed IVC-LEACH, a new hierarchical clustering protocol that tries to solve many of the shortcomings of the original Leach protocol.

IVC-LEACH provides numerous enhancements over the original Leach protocol, such as taking the value of the node into account when picking the cluster head and achieving load balance in power consumption across nodes during message transactions between nodes and the base. Furthermore, this technique does not rely on a random mechanism that results in the selection of an ineffectual cluster head. To ensure the IVC-LEACH protocol's effectiveness, it was simulated using Matlab and the simulation results were compared between the original LEACH protocol and the IVC-LEACH protocol.

The simulation findings of the original LEACH protocol revealed that the first node's power is completely depleted in round 964, and all nodes lose their entire capacity following completion of the 1491 role in the network. Although the first node dies after about 563 rounds in IVC-LEACH, the slope of the curve follows a uniform pattern, resulting in all nodes dying after around 2465 turns. As a result, the network's lifetime using the IVC-LEACH protocol is 50% more than the network's lifetime using the LEACH protocol.

This result demonstrates the success of the proposed IVC-LEACH protocol in accomplishing the research objectives of balancing power consumption between nodes and prolonging network life by 50% when compared to conventional LEACH, as well as increasing network stability and performance.

**Notes on contributors**

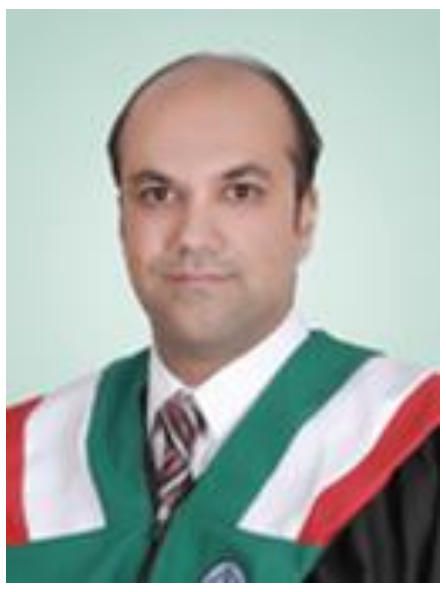

***Ghassan Samara*** *Holds BSc. and MSc. in Computer Science, and PhD in Computer Networks. He obtained his PhD, from Universiti Sains Malaysia (USM) in 2012. His field of specialization is Cryptography, Authentication, Computer Networks, Computer Data and Network Security, Wireless Networks, Vehicular Networks, Inter-vehicle Networks, Car to Car Communication, Certificates, Certificate Revocation, QoS, Emergency Safety Systems. Currently, Dr. Samara is an associate professor at Zarqa University, Jordan.*

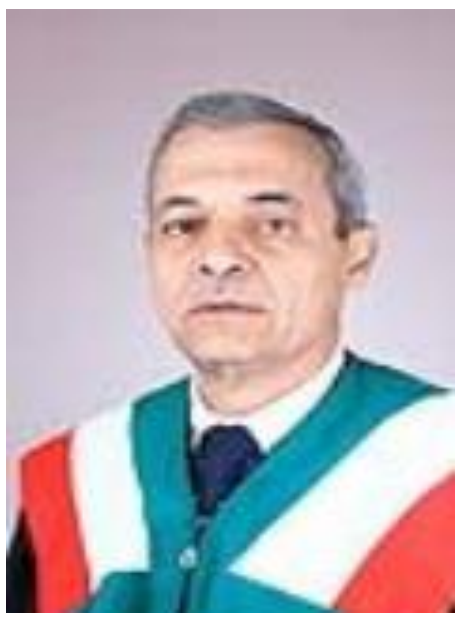

**Mohammad Hassan** received his BS degree from Yarmouk University in Jordan in 1987, the MS degree from Univ. of Jordan, in 1996, and the PhD degree in computer information systems from Bradford University, UK in 2003. He is working as an associate professor in the department of computer science at Zarqa University in Jordan. His research interest includes information retrieval systems and database systems.

***Yahya Zayed*** *Holds BSc. and MSc. in Computer Science from Zarqa University, Jordan.*